\documentclass[prd,preprint,nofootinbib]{revtex4-1} 

\usepackage{latexsym,verbatim}
\usepackage{amssymb}
\usepackage{amsfonts}
\usepackage{amsmath,color}
\usepackage[dvips]{graphicx}
\usepackage{bm}
\usepackage{soul}
\usepackage[utf8]{inputenc}
\usepackage[T1]{fontenc}

\usepackage{hyperref}

\hypersetup{
    colorlinks=true,
    linkcolor=red,
    citecolor=red,
    filecolor=magenta,      
    urlcolor=cyan,
    pdftitle={Physics is simple...},
    pdfpagemode=FullScreen,}

\def\ber{\begin{eqnarray}}
\def\eer{\end{eqnarray}}
\def\beq{\begin{equation}}
\def\eeq{\end{equation}}

\def\ed{\end{document}}

\def\dT#1{\frac{\mathrm{d} #1}{\mathrm{d}T}}
\def\DT#1{\frac{\mathrm{D} #1}{\mathrm{d}T}}

\def\DTO#1{\frac{\mathrm{D^{\bot}} #1}{\mathrm{d}T}}
\def\DL#1{\frac{\mathrm{D} #1}{\mathrm{d}\lambda}}
\def\DLFW#1{\frac{\mathrm{D_{{fw}}} #1}{\mathrm{d}\lambda}}

\def\dt#1{\frac{\mathrm{d} #1}{\mathrm{d}t}}

\def\dtau#1{\frac{\mathrm{d} #1}{\mathrm{d}\tau}}

\def\Dtau#1{\frac{\mathrm{D} #1}{\mathrm{d}\tau}}
\def\DtauU#1{\frac{\mathrm{D} #1}{\mathrm{d}\tau_{U}}}
\def\DtauO#1{\frac{\mathrm{D^{\bot}} #1}{\mathrm{d}\tau_{U}}}

\makeatletter
\let\jnl@style=\rm
\def\ref@jnl#1{{\jnl@style#1}}

\def\aj{\ref@jnl{AJ}}                   
\def\actaa{\ref@jnl{Acta Astron.}}      
\def\araa{\ref@jnl{ARA\&A}}             
\def\apj{\ref@jnl{ApJ}}                 
\def\apjl{\ref@jnl{ApJ}}                
\def\apjs{\ref@jnl{ApJS}}               
\def\ao{\ref@jnl{Appl.~Opt.}}           
\def\apss{\ref@jnl{Ap\&SS}}             
\def\aap{\ref@jnl{A\&A}}                
\def\aapr{\ref@jnl{A\&A~Rev.}}          
\def\aaps{\ref@jnl{A\&AS}}              
\def\azh{\ref@jnl{AZh}}                 
\def\baas{\ref@jnl{BAAS}}               
\def\bac{\ref@jnl{Bull. astr. Inst. Czechosl.}}
\def\caa{\ref@jnl{Chinese Astron. Astrophys.}}
\def\cjaa{\ref@jnl{Chinese J. Astron. Astrophys.}}
\def\icarus{\ref@jnl{Icarus}}           
\def\jcap{\ref@jnl{J. Cosmology Astropart. Phys.}}
\def\jrasc{\ref@jnl{JRASC}}             
\def\memras{\ref@jnl{MmRAS}}            
\def\mnras{\ref@jnl{MNRAS}}             
\def\na{\ref@jnl{New A}}                
\def\nar{\ref@jnl{New A Rev.}}          
\def\pra{\ref@jnl{Phys.~Rev.~A}}        
\def\prb{\ref@jnl{Phys.~Rev.~B}}        
\def\prc{\ref@jnl{Phys.~Rev.~C}}        
\def\prd{\ref@jnl{Phys.~Rev.~D}}        
\def\pre{\ref@jnl{Phys.~Rev.~E}}        
\def\prl{\ref@jnl{Phys.~Rev.~Lett.}}    
\def\pasa{\ref@jnl{PASA}}               
\def\pasp{\ref@jnl{PASP}}               
\def\pasj{\ref@jnl{PASJ}}               
\def\rmxaa{\ref@jnl{Rev. Mexicana Astron. Astrofis.}}%
\def\qjras{\ref@jnl{QJRAS}}             
\def\skytel{\ref@jnl{S\&T}}             
\def\solphys{\ref@jnl{Sol.~Phys.}}      
\def\sovast{\ref@jnl{Soviet~Ast.}}      
\def\ssr{\ref@jnl{Space~Sci.~Rev.}}     
\def\zap{\ref@jnl{ZAp}}                 
\def\nat{\ref@jnl{Nature}}              
\def\iaucirc{\ref@jnl{IAU~Circ.}}       
\def\aplett{\ref@jnl{Astrophys.~Lett.}} 
\def\apspr{\ref@jnl{Astrophys.~Space~Phys.~Res.}}
\def\bain{\ref@jnl{Bull.~Astron.~Inst.~Netherlands}}
\def\fcp{\ref@jnl{Fund.~Cosmic~Phys.}}  
\def\gca{\ref@jnl{Geochim.~Cosmochim.~Acta}}   
\def\grl{\ref@jnl{Geophys.~Res.~Lett.}} 
\def\jcp{\ref@jnl{J.~Chem.~Phys.}}      
\def\jgr{\ref@jnl{J.~Geophys.~Res.}}    
\def\jqsrt{\ref@jnl{J.~Quant.~Spec.~Radiat.~Transf.}}
\def\memsai{\ref@jnl{Mem.~Soc.~Astron.~Italiana}}
\def\nphysa{\ref@jnl{Nucl.~Phys.~A}}   
\def\physrep{\ref@jnl{Phys.~Rep.}}   
\def\physscr{\ref@jnl{Phys.~Scr}}   
\def\planss{\ref@jnl{Planet.~Space~Sci.}}   
\def\procspie{\ref@jnl{Proc.~SPIE}}   

\makeatother

\begin{document}

\author{Matteo Luca Ruggiero}
\email{matteoluca.ruggiero@unito.it}
\affiliation{Dipartimento di Matematica ``G.Peano'', Universit\`a degli studi di Torino, Via Carlo Alberto 10, 10123 Torino, Italy}
\affiliation{INFN - LNL , Viale dell'Universit\`a 2, 35020 Legnaro (PD), Italy}

\date{\today}

\title{Physics is simple only when analysed locally}

\begin{abstract}
The definition of a reference frame in General Relativity {is achieved through}  the construction of a congruence of time-like world-lines. In this framework, splitting techniques {enable us} to express physical phenomena in analogy with Special Relativity, {thereby} realizing the local description in terms of Minkowski spacetime in {accordance with}  the equivalence principle. This approach holds promise for elucidating the foundational principles of relativistic gravitational physics, as it illustrates how its 4-dimensional mathematical model manifests in practical measurement processes conducted in both space and time. In addition, we show how, within this framework, the Newtonian gravitational {force} naturally {emerges} as an effect of the {non-geodesic} path of the reference frame.
\end{abstract}

\maketitle


\section{Introduction}\label{sec:intr}

General Relativity (GR) is currently the most successful model we have for understanding gravitational interactions. Since its publication over 100 years ago, Einstein's theory has passed numerous observational tests, ranging from the space around Earth to the far reaches of the Universe \cite{2014arXiv1409.7871W, will2018theory}. However, there are still unresolved issues with the consequences of applying GR to large-scale structures, such as the problems of dark matter and dark energy \cite{universe2040023} and, in addition, we still do not know how to reconcile it with quantum mechanics. The latter, whose development culminated in the Standard Model of particle physics and its experimental success, is based on the belief that the ultimate nature of matter is  made of \textit{discrete}  entities, while  Einstein's theory suggests that gravity is the geometrical structure underlying the spacetime \textit{continuum}.

The concept of spacetime is indeed revolutionary with respect to classical physics, which is based on \textit{absolute time and space}: in Newton's words \cite{newton1850newton}, ``absolute, true and mathematical time, of itself, and from its own nature flows equably without regard to anything external[...]'' and ``absolute space, in its own nature, without regard to anything external, remains always similar and immovable.'' Accordingly,  space and time are independent of physical events and can be thought of as the stage or the arena where physical phenomena develop.
 From a physical-mathematical point of view, a geometric theory of Newtonian space and time can be obtained (see e.g. \citet{friedman2014foundations}), but it is deeply different from the relativistic framework, where only the union of space and time, the \textit{spacetime}, has a meaning. It is always a good thing to quote Minkowski words,  according to which  \cite{mink} ``[...] space by itself and time by itself are doomed to fade away into mere shadows, and only a kind of union of the two will preserve an independent reality.''

Spacetime is the mathematical model of  GR:  more precisely, it is a {(pseudo)Riemannian} manifold $%
\mathcal{M}^{}$, that is a pair $\left(
\mathcal{M},{g}\right)$, where $\mathcal{M}$ is a connected 4-dimensional Haussdorf manifold and $%
{g}$ is the metric tensor. The {Riemannian} structure implies that $\mathcal{M}$ is endowed
with an affine connection compatible with the metric, i.e. the
standard Levi-Civita connection; Special Relativity (SR)  is indeed a particular case, where  the model is a flat 4-dimensional manifold.  

If concepts are clear and evident from a mathematical perspective, the introduction of spacetime may appear to complicate matters from a common-sense viewpoint, as it cannot be visualized. Indeed, while the 4-dimensional Einsteinian approach has proven successful in understanding the structure of our Universe and gravitational phenomena, in our daily lives and laboratories we can only experience time (measured by clocks) and space (measured by rulers) as independent physical quantities. Accordingly, from an operational perspective and in order to deepen our understanding of the foundations of the theory, it is essential to determine which space and time measurements a given observer can perform in curved spacetime.

Thus, the question we face is how the theoretical framework of GR translates into actual measurements. Let us consider the case of SR and suppose we are in an inertial frame (IF), which we know is privileged for the formulation of physics laws.  Measurements operations require the existence of (synchronized) \textit{clocks} and \textit{rulers}, and these operations are {performed} by an \textit{observer}, whose  \textit{laboratory} has a definite spatial volume, and where measurements {take a finite amount of time}: differently speaking every measurement is necessarily local, since the {\textit{measurement domain}} is a reasonably small region of spacetime.  Additionally, we know how measurements change when we move to another IF: it is sufficient to apply the {Lorentz transformation.} 

It is useful to remark that the essence of Einstein's novelty {is not} so much about us living in a 4-dimensional Universe or introducing a mysterious fourth dimension. Even in Newtonian physics, motion is described naturally in four dimensions, considering three dimensions of space and one of time. What sets Einstein's theory apart is the recognition that space and time {are not} independent entities. This becomes evident when we employ the {Lorentz transformation}  between different inertial frames.

As we know, inertial frames continue to play a privileged role in SR as in Newtonian physics. On the other hand, we do not have ``privileged'' observers in GR, so we need to define with great care what is measured by observers in this case. Or, more simply, we need to define what is measured by accelerated observers in SR. This is, for instance, the case of a rotating frame: from one hand, this is important from an historical perspective, since Einstein used it as an example to relate acceleration to non-Euclidean geometry\cite{evolution} (this is the so-called Ehrenfest's paradox \cite{Rizzi:2002sk}); on the other hand, it is relevant since we are living on a rotating Earth. 

The series of procedures that, starting from a 4-dimensional mathematical model, allows one to attribute (in an operationally well-defined and unambiguous way) to a given observer the measurements that he/she can perform (along his/her world-line) is called \textit{spacetime splitting.} The basic idea is to project 4-dimensional quantities and equations along the world-line of the observer to obtain space-like quantities and time-like ones: in doing so, the measurement process is locally defined in analogy to what happens in SR. The development in tools for spacetime splitting have undergone significant, albeit diverse, progress over the years, finding application in various contexts within GR: the common
goal across different approaches is to focus on what a test family of observers, moving along certain curves in the 4-dimensional continuum, observes.  Multiple approaches to spacetime splitting exist, and recent efforts have been dedicated to unifying these methodologies within a cohesive framework, emphasizing the interconnections among the various techniques (see e.g. \cite{Jantzen:1992rg,Jantzen:1996au,RR2004s,de2010classical,Ruggiero:2023ker} and references therein). {In particular, we follow the notation introduced by \citet{Jantzen:1992rg} closely.}

In this paper we introduce a basic approach to spacetime splitting which, starting from the definition of reference frame based on the congruence of world-lines, {enables us} to understand how, along these  trajectories, measurements of space and time can be reconstructed, facilitating the translation of physical phenomena from SR into GR. In addition, in the description of the dynamics relative to the observer's reference frame, we show how the gravitational field, which is the geometry of spacetime in a relativistic framework, naturally emerges as a Newtonian gravitational force. 

We are of the opinion that this approach holds potential in emphasizing the foundations of relativistic gravitational physics, since it shows how its 4-dimensional mathematical model translates into actual measurement processes which are performed in space and time, thus avoiding the potential confusion arising from use of 4-dimensional equations and objects and their physical contents, which depend on the observer making the measurements: this is a natural way to understand the {meaning of relativity.} This approach may also prove effective   in teaching, since drawing parallels with the more familiar concepts of SR can also facilitate students' understanding of new GR ideas.

The paper is organized as follows: we focus on reference frames in Section \ref{sec:rf}, while in Section \ref{sec:proj} we introduce the projection technique which we apply to kinematical quantities in Section \ref{sec:kin}. Then, in Section \ref{sec:prop} we  analyze the properties of the congruence; the latter have an impact {on} the description of the dynamics, which is the subject of Section \ref{sec:reldesc}. Discussion and final remarks are in Section \ref{sec:remarks}.

\section{Reference Frames}\label{sec:rf}

\begin{figure}[t]
\begin{center}
\includegraphics[scale=.5]{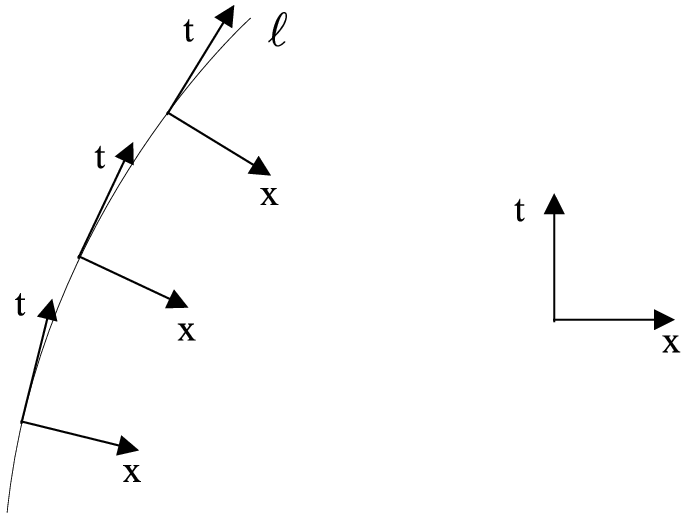}
\caption{ On the right, a reference frame in SR; on the left, along a world-line {$\ell$} at each event we can define a (local) reference frame.} \label{fig:sistema_rs_rg}
\end{center}
\end{figure}

To begin with, let us focus on the ideas of time and space  drawn from everyday experience.\\

\noindent \textbf{Time} is measured by a clock; in particular, every observer is provided with a clock which measures his/her proper time along his/her world line.\\

\noindent \textbf{Space} is made of an ensemble of places, a region where all clocks measure the same time: this fact is very important for a sharp division between time and space. \\

In SR time and space have a well defined meaning within an IF, but their measurement changes when moving from an IF to another: the link is provided by the {Lorentz transformation}. Things are different in GR, where do not exist a class of privileged observers: in other words, differently from SR, we do not have something similar to an inertial observer who can determine measurements performed by other observers, at rest with respect to him/her. So we lose the possibility to have global observers in GR. 

A natural solution comes when we remember that ``physics is simple only when analyzed locally''\cite{MTW}: accordingly, we have to abandon the ambitious plan to distinguish space and time \textit{globally}, i.e. in the whole manifold, and we must be content  to split spacetime into space and time \textit{locally}, which means around a given event along the world-line of a reference observer (see Figure \ref{fig:sistema_rs_rg}). Starting from these ideas, we are going to briefly review the concept of reference frame in Newtonian physics and SR, to introduce its natural extension in GR.

The definition of reference frame is preliminary to the study of mechanics: the intuitive idea of motion requires the definition of reference points that are at rest, so that it becomes meaningful to describe a change of position with respect to them. In Newtonian physics reference frames   are defined by a set of  reference points which are at rest with respect to each other, so that we can speak of a \textit{rigid} frame; these points are physically identified by ideal markers which, in turn, {enable them to be mathematically identified using suitable coordinates.} Within a Newtonian reference frame, each point is ideally provided with a clock, and the whole set of clocks  are synchronized to perform measurements of kinematical quantities. We remark that  the construction of a reference frame is meaningful only in terms of measurements, which are typically performed by an observer or a device which is at rest in the frame; additionally, { measurements are significant only when they are referred to an observer who can interpret them in terms of coordinate-independent quantities.}

In Newtonian physics measurements of space and time are absolute and therefore independent of the reference system in which they are performed. On the contrary, in a relativistic context, we understand that the outcome of space and time measurements varies for observers in motion relative to each other.  More generally speaking, measurements in a relativistic framework are inherently dependent on the observer, thus necessitating a criterion for comparing measurements made by different observers. A physical observer comprises a collection of measuring instruments, such as clocks, rulers and so on. In addition, it is important to emphasize that  these instruments must operate within a limited space (the observer's laboratory) and over a limited time interval to disregard curvature effects, allowing for an unambiguous division of spacetime into space and time, as per the principle of equivalence.

The concept of \textit{congruence} {allows us} to naturally introduce a relativistic definition of reference frame. 

A \textit{physical reference frame} is defined as a time-like {congruence} $\mathcal C$: the latter is a collection of the world-lines of test particles that constitute the \textit{reference fluid}; these test particles have the same role as the reference points of Newtonian physics.  The term ``congruence'' refers to a set of world-lines that fill the manifold or a portion of it, smoothly, continuously, and without intersecting; they are time-like since they are the spacetime trajectories of material particles, which means that at each event along them the tangent vector is time-like.  

The idea of a reference fluid extends naturally from the concept of a reference solid, which, as we said, is  meaningful in Newtonian physics and applicable in flat spacetime where test particles form a global inertial frame. In such cases, the relative distances between particles remain constant, resembling the behavior of a rigid frame. However in GR test particles may experience gravitational fields (curvature of spacetime), while in SR they may encounter acceleration fields.  In both scenarios, {global inertiality} is compromised, leading to effects that cause variations in {the distances} between particles. It therefore becomes necessary to refer to a reference fluid, renouncing the requirement of classical rigidity.

A congruence $\mathcal C$ is characterized by the field of unit vectors $u$ tangent to its world-lines parameterized with the proper time, so that we have\footnote{We use geometrical units such that $c=G=1$ here and henceforth; spacetime signature is $(-,+,+,+)$.}
\beq
u \cdot u =-1. \label{eq:normu1}
\eeq
In order the emphasize the  dependence on $u$, we will denote the congruence by $\mathcal C_{u}$.  Accordingly, the integral curves of $u$ are the world-lines of the particles which constitute the reference frame. In physical terms,  an  observer therefore is to be defined by a (narrow) congruence of time-like world-lines representing the points at fixed positions in his/her laboratory.

\subsection{Reference Frames in Flat Spacetime}\label{ssec:flatrf}

\begin{figure}[t]
\begin{center}
\includegraphics[scale=.8]{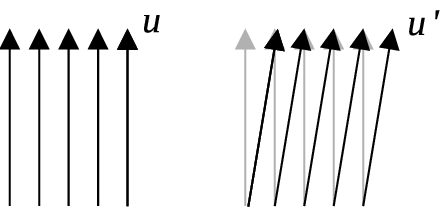}
\caption{The  reference frame $\mathcal C_{u}$ is defined by the integral curves of the vector field $u$; the reference frame $\mathcal C_{u'}$ is defined by the integral curves of $u'$.} \label{fig:campi_u0}
\end{center}
\end{figure}

Let us apply the above definition to the simple case of an inertial reference frame in Minkowski spacetime:  in particular, we may say that a reference frame defined by $\mathcal C_{u}$ is \textit{inertial} if the vector field $u$ is constant. In Figure \ref{fig:campi_u0} we see the reference frame $\mathcal C_{u}$ defined by the integral curves of the constant field $u$: each of them represents the world-line of a particle of the reference frame, and they are at rest with respect to each other; the vector field $u'$ defines another reference frame, $\mathcal C_{u'}$, whose particles are in motion with respect  to $\mathcal C_{u}$, with constant velocity. Starting from $u$ we can define a Cartesian coordinate basis in the given inertial reference frame. To do this, we introduce a time coordinate $t$ such that $u=\partial_{t}$; in other words the curves $t =\mathrm{var}$ are the integral curves of $u$. Then, we consider three space-like orthogonal directions, and introduce the corresponding tangent coordinate basis vectors $\partial_{x},\partial_{y},\partial_{z}$; the hypersurfaces spanned by these vectors are orthogonal to the time direction, so their points share the same time coordinate.

  In summary,
the spacetime metric $\eta_{\mu\nu}$ is defined by the conditions (scalar products of the basis vectors)
\beq
\eta_{\alpha\beta}=\partial_{\alpha} \cdot \partial _{\beta}, \label{eq:normeta1}
\eeq
where $\alpha,\beta=t,x,y,z$; hence the metric takes the form
{
\beq
\eta_{\mu\nu}=\left(\begin{array}{cccc}-1{} & 0 & 0 & 0 \\0 & 1 & 0 & 0 \\0 & 0 & 1 & 0 \\0 & 0 & 0 & 1\end{array}\right).
\eeq}

Given an origin event $O$ and a basis $\partial_{t},\partial_{x},\partial_{y},\partial_{z}$, an event $P$ is identified by the 4-vector
\beq
OP = X^{\mu}\partial_{\mu}= t\, \partial_{t}+x\,\partial_{x}+y\,\partial_{y}+z\,\partial_{z}
\eeq
whose components with respect to the coordinates basis are $X^{\mu}=(t,x,y,z)$.
 A set of Cartesian orthonormal coordinates induced by the basis $\partial_{t},\partial_{x},\partial_{y},\partial_{z}$ with origin $O$ is called \textit{Cartesian reference system} adapted to $\mathcal C_{u}$: the corresponding coordinates $\left(t,x,y,z \right)$ are said to be \textit{adapted to the reference frame.}

The world-line  of a particle is a curve  ($C^{1}$ at least) in spacetime; in particular, it can be represented in the form $OP= R(\tau)$, where $\tau$ is the proper time, i.e. the time measured by a clock carried by the particle.  Accordingly, we may define 
\beq
 V=\dtau{ R}, \label{eq:defV}
\eeq
which is the 4-velocity, or ``absolute velocity'', i.e. the vector tangent to the curve. We can give the following \textit{relative  formulation} of the position vector $OP$
\beq
 R=  r+t u, \label{eq:defRrel}
\eeq
where
\beq
r=x\,\partial_{x}+y\,\partial_{y}+z\,\partial_{z}  \label{eq:defrrel} 
\eeq
defines the \textit{relative position}, and $t$ the \textit{relative time}. It is important to emphasize that the word \textit{relative} refers to the fact that both position and time are relative to the given reference frame, and they can generally change when we consider a different reference frame.

From the definition (\ref{eq:defV}) we may write
\beq
 V=\dtau{ R}=\dt{ R}\dtau{t}=\dtau{t}\left(\dt{ r} +u \right), \label{eq:defV1} 
\eeq
or
\beq
V=\gamma\left( v+u \right), \label{eq:defV2} 
\eeq
where
\beq
 v=\dt{ r} \label{eq:defv}
\eeq
is the \textit{relative velocity} of the particle, and 
\beq
\gamma=\dtau{t}. \label{eq:defgamma}
\eeq
Since $u \cdot v=0$, from (\ref{eq:defV2}) we obtain
\beq
\gamma=-{ V \cdot u}, \label{eq:defgamma2}
\eeq
and, using the fact that $V^{2}=-1$ (it is an invariant, and it is $v=0$ in the reference frame where the particle is at rest), it is possible to write the \textit{Lorentz factor} $\gamma$ in the usual form
\beq
\gamma=\frac{1}{\sqrt{1-{v^{2}}{}}}. \label{eq:defgamma3}
\eeq

\subsection{Reference Frames in Curved Spacetime}\label{ssec:curvedrf}

\begin{figure}[t]
\begin{center}
\includegraphics[scale=.8]{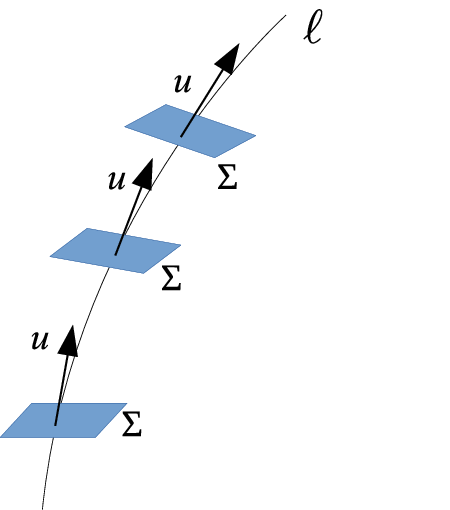}
\caption{At each point along {$\ell$}, $\Sigma$ is the 3-dimensional space orthogonal to $u$.} \label{fig:sistema_rg_sigma0}
\end{center}
\end{figure}

The above description shows how an inertial frame is defined in flat spacetime in terms of a congruence $\mathcal C_{u}$, where $u$ is a constant field; in this case, the congruence induces a global splitting of spacetime in terms of a time direction, defined by the world-line whose tangent vector is $u$, and a 3-dimensional space, which is globally defined by directions whose tangent vectors are orthogonal to $u$; the events in the three space are simultaneous. This construction is relative to $u$, but it is globally defined in the whole spacetime: as a matter of fact, different observers induce a different splitting and, in particular, different hypersurfaces of simultaneity.

A similar approach can be used to define a splitting procedure in curved spacetime or, also, for accelerated observers in flat spacetime (thus relaxing the constraint of constancy of $u$).  To this end, let {$\ell \in \mathcal C_{u}$} be the reference world-line, which can be {thought} of as the world-line of the reference observer; this is a generic time-like curve in spacetime (not necessarily a straight line, as for inertial observers in flat spacetime). Then, we denote with $\Sigma$ the 3-dimensional space orthogonal to  $u$, at each point along {$\ell$}: $\Sigma$ is called \textit{local rest space, or local space  platform}; while $u$ evolves along {$\ell$}, $\Sigma$ describes a volume that is the evolution of the laboratory (see Figure \ref{fig:sistema_rg_sigma0}). 

{The structure defined by $u \oplus \Sigma$ is the basis of the measurement process, since it locally reproduces what happens in SR: we have a time direction and 3-dimensional space orthogonal to it. This is a somewhat natural way to define a reference system in curved spacetime and it constitutes what is called \textit{non-linear} or generalized reference system.} 

{Let $\{x^{\mu }\}=(x^{{0}},x^{1},x^{2},x^{3})$ be a system of
coordinates\footnote{Latin indices run from 1 to 3, and refer to space components, while Greek indices run from 0 to 3, and label spacetime components.} in the neighborhood of a point $p\in \mathcal{M}$; 
these coordinates are said to be \textit{admissible} with respect
to the congruence $\mathcal C_{u}$ if the following conditions are satisfied\footnote{Note that this is a consequence of the chosen spacetime signature. $(-,+,+,+)$.}:
\beq
g_{{00}}<0\ \ \ \ g_{ij }dx^{i }dx^{j }>0, \label{eq:adms}
\eeq
where $g_{\mu\nu}$ are the components of the metric tensor (here and henceforth, we use the approach introduced by \citet{cattaneo1958}). The coordinates lines $x^{{0}}=var$ can be interpreted as describing the world-lines of the particles of the reference fluid. In other words, when using a system of admissible coordinates, the lines $x^{0}=var$ coincide with the integral curves of $u$. Physically, this means that the observer is at rest in this coordinate system. Note that this naturally generalizes to curved spacetime the definition of adapted coordinates introduced in Section \ref{ssec:flatrf}. For each particle in the reference fluid, the time variable $x^{0}$ can be interpreted as the reading of a clock attached to the particle, operating in an arbitrary manner, subject only to the condition that $\displaystyle \dtau{x^{0}}>0$ is positive, where $\tau$ is the proper time  for the particle under consideration. Such a clock is referred to as a coordinate clock, and the time 
it measures is known as coordinate time. While the coordinate time is naturally defined in this way along the world-lines of the congruence, the spatial directions are defined (within $\Sigma$) up to rotations; the events in $\Sigma$ are simultaneous. That being said,  we can introduce a coordinate basis $\partial_{\mu}, \mu=0,1,2,3$, with $\partial_{0}=u$.} 

{As we will discuss in Section \ref{sec:prop}, the properties of the congruence impact the measurement process: these properties are determined by the geometry of the vector field $u$, which has a coordinate-independent meaning. However, in many cases, even though it is not a mandatory choice, the use of adapted coordinates allows for better physical insight. Therefore, we conclude this Section by examining in more detail the consequences of the choice of adapted coordinates.}




Let $u^\mu$ be the components of the field of unit vectors
 tangent to the world-lines of the congruence
$\mathcal C_{u}$ in adapted coordinates, parameterized by $\tau_{u}$, i.e. the proper time, i.e.
\beq
u^{\mu}=\frac{dx^\mu}{d\tau_{u}} \label{eq:defumu1}.
\eeq
The spatial coordinates $x^i$ are constant
along the lines of $\mathcal C_{u} $ and  $dx^i=0$ along any
line of $\mathcal C_{u} $; the same holds for the components
$\displaystyle u^{i}=\frac{dx^i}{d\tau_{u}}=0$. Hence, the
component $u^0$ directly comes from the condition
\beq
u\cdot u= -1 \leftrightarrow g_{\mu \nu
}u ^{\mu }u ^{\nu }=-1, \label{eq:normug}
\eeq
where $g_{\alpha\beta}$ is the spacetime metric in the given coordinates;
in fact, we get
\begin{equation}
u ^{{0}}=\frac{1}{\sqrt{-g_{{00}}}}. \label{eq:gammazeroup}
\end{equation}
The covariant components can be found in the usual way
\begin{equation}
u_0=g_{0\mu}u^\mu=g_{00}u ^{0}=-\sqrt{-g_{00}},
\label{eq:gammazerodown}
\end{equation}
\begin{equation}
u_{i}=g_{i\mu}u^\mu=g_{i0}u^0. \label{eq:gammai}
\end{equation}
In conclusion, we have
\begin{equation}
\left\{
\begin{array}{l}
u ^{{0}}=\frac{1}{\sqrt{-g_{{00}}}}, \\
u ^{i}=0,
\end{array}
\right. \;\;\;\;\;\;\;\;\;\;\;\;\left\{
\begin{array}{l}
u _{0}=-\sqrt{-g_{{00}}}, \\
u _{i}=g_{i{0}}u ^{{0}}.
\end{array}
\right.  \label{eq:gammas11}
\end{equation}\\
{The reference system defined by $\mathcal C_{u}$ is said to be \textit{time-orthogonal}  when there exists at least one adapted chart in which $g_{0i}=0$. In this case,  the coordinate lines $x^{0} = var$ are orthogonal to the 3-manifold $x^{0} = const$:  consequently, we observe that $u_{i}=0$, according to Eq. (\ref{eq:gammas11}). This is indeed a geometric property of the congruence and, as such, within the reference frame, it does not depend on the coordinates we use: as we will see in Section \ref{sec:prop}, this condition corresponds to the case  where the \textit{vorticity} of the vector field $u$ is null.}

Summarizing: in flat spacetime  it is possible to define a unique family of observers in the whole spacetime, i.e. a global inertial reference frame, and this is characterized by a congruence $\mathcal C_{u}$, where $u$ is a constant vector field. The Lorentz  transformation enables us to understand how to relate the measurements made in different inertial reference frames.

On the other hand, in curved spacetime, we don't have a natural family of privileged observers, so we can treat all of them on equal footing (even if, as we will discuss, some of them are more equal than others). Starting from the congruence, we can locally build a reference system and understand how measurements are made within it. In particular, the same approach can be applied to non-inertial observers in SR.

In the following Sections we will describe how it is possible to relate 4-dimensional ``entities'' (such as 4-vectors, tensors, and so on) to measurable quantities along the congruence. To clarify, we will describe the measurements of kinematical quantities and the description of the dynamics as observed by the reference observers

\section{The Projection Technique}\label{sec:proj}

The construction of a reference frame is based on the definition of congruence; accordingly, the 4-dimensional spacetime can be split {into} $u \oplus \Sigma$, and this splitting is the basis of the measurement process, since {it allows for the natural projection} of 4-dimensional entities along the time direction parallel to $u$, and onto $\Sigma$. Formally, this is done by using the \textit{orthogonal projection operators or projectors} that we are going to discuss.

Let $u^{\alpha}$ the components of the vector field $u$ defining the congruence $\mathcal C_{u}$, which satisfies the normalization condition (\ref{eq:normug}) in terms of the components of the spacetime metric $g_{\alpha\beta}$. We define the \textit{time projector}
\beq
T(u)^{\alpha}_{\beta}=-u^{\alpha}u_{\beta}, \label{eq:defT1}
\eeq
and the \textit{space projector}
\beq
P(u)^{\alpha}_{\beta}=\delta^{\alpha}_{\beta}+u^{\alpha}u_{\beta}. \label{eq:defP1}
\eeq
Clearly, it is  $\delta^{\alpha}_{\beta}=T(u)^{\alpha}_{\beta}+P(u)^{\alpha}_{\beta}$; additionally, we remark that the very definitions of these projectors depend on the vector field $u$. It is easy to check that they are projectors, i.e.
\begin{eqnarray}
 P(u)^{\alpha}_{\beta}P(u)^{\beta}_{\gamma}&=&P^{\alpha}_{\gamma}(u), \label{eq:proj1} \\
 T(u)^{\alpha}_{\beta}T(u)^{\beta}_{\gamma}&=&T^{\alpha}_{\gamma}(u), \label{eq:proj2} \\
 P(u)^{\alpha}_{\beta}T(u)^{\beta}_{\gamma}&=&T^{\alpha}_{\beta}(u)P^{\beta}_{\gamma}(u)=0. \label{eq:proj3}
\end{eqnarray}
In addition, we may write
\beq
g_{\alpha \beta}=T(u)_{\alpha\beta}+P(u)_{\alpha\beta}, \quad g^{\alpha\beta}=T(u)^{\alpha\beta}+P(u)^{\alpha\beta}. \label{eq:projmetric1}
\eeq

Let us consider a vector field $X$ with components $X^{\alpha}$, and project it with $T(u), P(u)$. We obtain
\beq
T^{\beta}_{\alpha}(u)X^{\alpha}=-u^{\beta}(u_{\alpha}X^{\alpha}), \label{eq:projXT}
\eeq
\beq
P^{\beta}_{\alpha}(u)X^{\alpha}=X^{\beta}+u^{\beta}(u_{\alpha}X^{\alpha}). \label{eq:projXP}
\eeq

Hence $X^{\alpha}=u^{\beta}(-u_{\alpha}X^{\alpha})+P^{\beta}_{\alpha}(u)X^{\alpha}$: then, we see that  its component along the $u$ direction is given by the scalar $-u_{\alpha}X^{\alpha}$, while the spatial projection  gives the 3-vector $P^{\beta}_{\alpha}(u)X^{\alpha}\in \Sigma$. The same can be done for an arbitrary rank tensor, by applying to each of its indices the projectors. For instance, if we consider the tensor $S^{\gamma}_{\beta}$, we obtain
\begin{align}
S^{\alpha}_{\epsilon}&=\left(T(u)^{\alpha}_{\gamma}+P(u)^{\alpha}_{\gamma}\right)\left(T(u)^{\beta}_{\epsilon}+P(u)^{\beta}_{\epsilon}\right)S^{\gamma}_{\beta}=u^{\alpha}u_{\epsilon}\left(u_{\gamma}u^{\beta}S^{\gamma}_{\beta} \right)+u^{\alpha}\left(-u_{\gamma}P(u)^{\beta}_{\epsilon}S^{\gamma_{\beta}} \right)+\\
& +u_{\epsilon}\left(u^{\beta} P^\alpha_{\gamma}(u)S^{\gamma}_{\beta}\right)+P(u)^{\alpha}_{\gamma}P(u)^{\beta}_{\epsilon}S^{\gamma}_{\beta}. \label{eq:projStensor}
\end{align}
In other words, the projection gives the scalar $u_{\gamma}u^{\beta}S^{\gamma}_{\beta} $, the 1-form $-u_{\gamma}P(u)^{\beta}_{\epsilon}S^\gamma_{\beta}$, the vector $u^{\beta} P(u)^\alpha_{\gamma}S^{\gamma}_{\beta}$ and the spatial tensor $P(u)^{\alpha}_{\gamma}P(u)^{\beta}_{\epsilon}S^{\gamma}_{\beta} $.\\

As a first application of the space and time projectors, we consider  two infinitesimally close spacetime events $P,P'$, in the vicinity of the reference world-line, such that their  separation in the given coordinates is $dx^{\alpha}$. The spacetime invariant between these two events is $ds^{2}=g_{\alpha\beta}dx^{\alpha}dx^{\beta}$. Using the first relation in Eq. (\ref{eq:projmetric1}), and setting 
\beq
\gamma_{\alpha\beta}\doteq P(u)_{\alpha\beta}=g_{\alpha\beta}+u_{\alpha}u_{\beta}, \label{eq:defgammaij}
\eeq
 we get
\beq
ds^{2}=T(u)_{\alpha\beta}dx^{\alpha\beta}+\gamma_{\alpha\beta}dx^{\alpha\beta} \leftrightarrow ds^{2}=-u_{\alpha}u_{\beta}dx^{\alpha}dx^{\beta}+\gamma_{\alpha\beta}dx^{\alpha}dx^{\beta},  \label{eq:orthdecomp2}
\eeq
or
\beq
ds^{2}=-dT^{2}+d\sigma^{2}, \label{eq:orthdecomp3}
\eeq
where $dT$ is the \textit{relative time}
\beq
dT=- u_{\alpha}dx^{\alpha} \label{eq:defreltime1}
\eeq
and 
\beq
d\sigma^{2}=\gamma_{\alpha\beta}dx^{\alpha}dx^{\beta} \label{eq:orthdecomp4}
\eeq
is the \textit{spatial distance} between the two events, i.e. the projection of their spacetime separation onto $\Sigma$. It is easy to check that in adapted coordinates the components of the space projector $P(u)_{\mu\nu} =  \gamma _{\mu \nu }$ are purely spatial, i.e. $\gamma_{\mu\nu}=\gamma_{ij}$: in particular, $\gamma_{ij}$ represents the spatial metric in $\Sigma$ (see below).

To better understand the meaning of Eq. (\ref{eq:orthdecomp3}), let us consider a couple of simple examples, in $1+1$ Minkowski spacetime, with coordinates $t,x$. To begin with, we take into account the congruence of inertial observers defined by $u=\partial_{t}$: in this case, it easy to check that $dT^{2}=dt^{2}$, and $d\sigma^{2}=dx^{2}$. Accordingly, the relative time is simply expressed by the coordinate time $t$, and space distance by $x$: differently speaking, these coordinates express time and space as measured by the reference observer.

Now, let us consider the congruence defined by $\displaystyle u=\gamma\left(\partial_{t}+v\partial_{x} \right)$ where $\displaystyle \gamma=\frac{1}{\sqrt{1-v^{2}}}$; this is again an inertial congruence of observers that are moving with constant speed $v$ in the reference frame to which the coordinates $t,x$ are adapted. In this case, we get
\beq
dT^{2}=u_{\alpha}dx^{\alpha}u_{\beta}dx^{\beta}=\left(dt-vdx \right)^{2}\gamma^{2} \label{eq:lorentz1}
\eeq
and
\beq
d\sigma^{2}=\gamma_{\alpha\beta} dx^{\alpha}dx^{\beta}=\left(dx-vdt \right)\gamma^{2}. \label{eq:lorentz2}
\eeq 

We see that the above Eqs. (\ref{eq:lorentz1}),(\ref{eq:lorentz2}) express the Lorentz  transformation: in fact, the relative time and spatial distance in the reference frame adapted to the congruence defined by $u$ are nothing but the coordinate time and space intervals measured by the observer identified by $u$.

More generally speaking, Eq. (\ref{eq:orthdecomp3}) expresses the orthogonal decomposition of the spacetime invariant into  $u \oplus \Sigma$. We see that this decomposition {allows us} to locally obtain a Minkowski structure, i.e. a spacetime splitting which naturally leads to space and time measurements.

In what follows we apply this approach to the measurement of kinematical quantities along the congruence.

\section{Projection of Kinematical Quantities}\label{sec:kin}

We consider a test particle  and we want to study its motion as seen from the reference frame defined by $\mathcal C_{u}$. {In terms of coordinates, its world-line is $x^{\beta}(\tau_{U})$, parameterized by the proper time $\tau_{U}$ measured along it.} First, we remark that if we suppose that this particle is moving with respect to the congruence $\mathcal C_{u}$ its world-line differs from any of the world-lines of the congruence. That being said, we suppose that in its spacetime evolution the particle moves from the event $P$ whose coordinates are $x^{\beta}$, to $P'$ whose coordinates are $x^{\beta}+dx^{\beta}$. We want to obtain the velocity that the observer attributes to the motion of {the} test particle.  To this end, {we first project} the infinitesimal spacetime $dx^{\beta}$ displacement onto $\Sigma$:
\beq
dx^{\beta} \rightarrow P(u)^{\alpha}_{\beta} dx^{\beta}, \label{eq:relvel1}
\eeq
to obtain the space displacement as measured by the observer. Then, we project  $dx^{\beta}$  along $u$
\beq
dx^{\beta} \rightarrow T(u)^{\alpha}_{\beta} dx^{\beta}=-u^{\alpha}u_{\beta}dx^{\beta}, \label{eq:relvel10}
\eeq
to get the relative time measured {by the observer:}
\beq
dT=-u_{\beta}dx^{\beta}. \label{eq:reltime1}
\eeq
{Note} that since both $u^{\beta}$ and $dx^{\beta}$ are time-like, the above quantity is always different from zero.

We are in position to calculate the \textit{relative velocity}, i.e. the velocity measured by the observer on the basis of his/her space and time measurements:
\beq
v^{\alpha}=\frac{ P(u)^{\alpha}_{\beta} dx^{\beta}}{dT}=\frac{\left(\delta^{\alpha}_{\beta}+u^{\alpha}u_{\beta}\right)dx^{\beta}}{-u_{\alpha}dx^{\alpha}} = \frac{\left(\delta^{\alpha}_{\beta}+u^{\alpha}u_{\beta}\right)U^{\beta}}{-u_{\alpha}U^{\alpha}}  \label{eq:relvel2}
\eeq
where we introduced the 4-velocity of the test particle $\displaystyle U^{\alpha}=\frac{dx^{\alpha}}{d\tau_{U}}$. 

Eq. (\ref{eq:relvel2}) can be rewritten as
\beq
v(U,u)^{\alpha}=\frac{U^{\alpha}+u^{\alpha}u_{\beta}U^{\beta}}{-u_{\alpha}U^{\alpha}}. \label{eq:relvel3}
\eeq
The notation adopted emphasizes the fact that $v(U,u)^{\alpha}$ is the velocity of the particle whose 4-velocity is $U$ with respect to the congruence defined by $u$. From Eq. (\ref{eq:relvel3}) we obtain
\beq
U^{\alpha}=\gamma(u,U)\left[u^{\alpha}+v(U,u)^{\alpha} \right], \label{eq:relvel4}
\eeq
where 
\beq
\gamma(U,u)\doteq-U_{\alpha}u^{\alpha}=\frac{1}{\sqrt{1-\beta^{2}}}, \label{eq:defgamma11}
\eeq
with $\beta^{2}=v_{\alpha}v^{\alpha}$.
We {note} that Eq. (\ref{eq:relvel4}) is the natural generalization in curved spacetime of Eq. (\ref{eq:defV2}), {which was} obtained in flat spacetime.

If $m$ is the rest mass of the test particle having  4-velocity $U$, then 
\beq
P^{\alpha}=m U^{\alpha} \label{eq:4mom1}
\eeq
represents its 4-momentum, to which we apply the space and time projectors. Using Eq. (\ref{eq:relvel4}), we then obtain
\beq
P(u)^{\beta}_{\alpha}P_{\beta}=m \gamma(U,u) v_{\alpha}(U,u) \doteq p_{\alpha}(U,u). \label{eq:4mom3}
\eeq
Accordingly, $p(U,u)$ is the spatial momentum of the particle relative to the observer $u$. If we apply the time projector, we get
\beq
T(u)^{\beta}_{\alpha}P_{\beta}=-u_{\alpha}u^{\beta}P_{\beta} =u_{\alpha}E(U,u), \label{eq:4mom4}
\eeq
where 
\beq
E(U,u)=-u^{\beta}P_{\beta}=-m u^{\beta} U_{\beta}=m \gamma(U,u) \label{eq:4mom5}
\eeq
is the relative energy of a particle with rest mass $m$: in fact, since in our notation  $c=1$, the above relation is equivalent to $E=m \gamma(U,u) c^{2}$. 

From Eq. (\ref{eq:4mom3}) we see that $|p(U,u)|=m\gamma |v|$.  The following relation holds true
\beq
m^{2}=E^{2}(U,u)-|p(U,u)|^{2}. \label{eq:4mom6}
\eeq
Accordingly, for a massless particle $m=0$ we obtain $E(U,u)=|p(U,u)|$, which is in agreement with the fact that $|v(U,u)|=1$.  So, we may write the natural splitting of the 4-momentum $P^{\alpha}$ in the form
\beq
P=E(P,u)u+p(P,u)=E(P,u)\left[u+v(U,u)\right]  \label{eq:4mom7}
\eeq

In summary, we have seen how it is possible to recover the local Minkwoski structure using the projection technique;  differently speaking, this approach naturally {enables us} to obtain from the 4-dimensional quantities $U^{\alpha},P^{\alpha}$ the relative quantities $v^{\alpha}(U,u), p^{\alpha}(P,u), E(P,u)$ which are measured along the congruence $C_{u}$.

Before ending this Section, we point out an important consequence of Eq. (\ref{eq:reltime1}). Let $\tau_{U}$ be the proper time measured along the world-line of the particle moving with 4-velocity $U$. Then, from (\ref{eq:reltime1}) we get
\beq
\frac{dT}{d\tau_{U}}=-u_{\beta}\frac{dx^{\beta}}{d\tau_{U}} = -u_{\beta}U^{\beta} = \gamma (u,U), \label{eq:reltime2}
\eeq
where we used Eq. (\ref{eq:defgamma11}). Accordingly, we have
\beq
\frac{dT}{d\tau_{U}}= \gamma (u,U) \label{eq:reltime3}
\eeq
which is the relation between the infinitesimal time interval $d T$ measured by the observer along the congruence and the corresponding proper time interval $d\tau_{U}$ measured along the particle world-line.

\section{Properties of the Congruence}\label{sec:prop}

The congruence is made of the world-lines constituting the reference fluid;  the properties of these world-lines have a relevant impact {on the description} of what is measured by the reference observers. For instance,  if we refer to the Einstein's elevator, we know that an observer in free fall does not feel the effect of the gravitational field: so, we do expect that something peculiar happens for congruences made of geodesic world-lines. 

{The characteristics of the motion of the particles of the reference fluid   are obtained  by applying the covariant derivative operator (denoted by a semicolon)} to the vector field $u$ (see e.g. \citet{poisson2004relativist,de2010classical} and references therein).  We remember that $u \cdot u =-1 \leftrightarrow u_{\alpha}u^{\alpha}=-1$: as a consequence
\beq
u_{\alpha}u^{\alpha}_{\,\,;\beta}=0. \label{eq:orthcovder1}
\eeq
Then, we can project $u^{\alpha}_{\,\,;\beta}$ as a $(^{1}_{1})$-type tensor (see Section \ref{sec:proj}) to obtain
\beq
u^{\alpha}_{\,\,;\beta}=-a(u)^{\alpha}u_{\beta}-k(u)^{\alpha}_{\,\,\beta} \label{eq:projcovder1}
\eeq
where $a(u)=\nabla_{u} u$ is the (covariant) acceleration\footnote{$a(u) \in \Sigma$, i.e. $a(u) \cdot u=0$ or $P(u) a(u)=a(u)$.} and 
\beq
k^{\alpha}_{\,\,\beta}=-P(u)^{\alpha}_{\epsilon}P(u)^{\gamma}_{\beta}u^{\epsilon}_{\,\,\gamma}. \label{eq:defkappa1}
\eeq
In particular, we may write the tensor $k$ in the form
\beq
k_{\alpha\beta}=-\theta_{\alpha\beta}+\omega_{\alpha\beta}, \label{eq:defkappa2}
\eeq
with
\beq
\theta_{\alpha\beta}=P(u)^{\gamma}_{\alpha}P(u)^{\delta}_{\beta} u_{(\gamma;\delta)}, \quad  \omega_{\alpha\beta}=-P(u)^{\gamma}_{\alpha}P(u)^{\delta}_{\beta} u_{[\gamma;\delta]},  \label{eq:defkappa3}
\eeq
where we introduced the symmetric part $\theta_{\alpha\beta}$, which is called \textit{expansion}, and the antisymmetric part $\omega_{\alpha\beta}$, which is called \textit{vorticity}: as discussed by \citet{poisson2004relativist}, the latter is connected  to the relative rotation of the world-lines of the reference fluid, while the former tells us how the world-lines are expanding. {In addition, when the vorticity is zero, the congruence is said to be \textit{hypersurface orthogonal}, which means that spacetime can be foliated by a family of spacelike hypersufaces: consequently, the set of comoving coordinates along the congruence is defined in the whole spacetime. }

Let us consider two congruences $\mathcal C_{u}$ and $\mathcal C_{U}$, and let $\tau_{U}$ the proper time along the vector field $U$. Then, we can calculate
\beq
\frac{\mathrm{D}u}{\mathrm{d}\tau_{U}} \doteq \nabla_{U}u.  \label{eq:covderU}
\eeq
We use Eq. (\ref{eq:relvel4}) and write $\displaystyle U^{\alpha}=\gamma(u,U)\left[u^{\alpha}+v(U,u)^{\alpha} \right]$; hence, substituting in Eq. (\ref{eq:covderU}) and using the properties of the covariant derivative operator, we obtain
\beq
\frac{\mathrm{D}u}{\mathrm{d}\tau_{U}} =\gamma \left[a(u)-k(u)\lfloor v(U,u) \right], \label{eq:covderUproj1}
\eeq
where $k(u)\lfloor v(U,u) \equiv k^{\alpha}_{\,\, \beta}v^{\beta}$.  

Let $\epsilon_{\alpha\beta\gamma\delta}$ be the completely antisymmetric symbol, from which we may define  the unit volume 4-form $\eta_{\alpha\beta\gamma\delta}=\sqrt{|g|}\epsilon_{\alpha\beta\gamma\delta}$. Then, if we set
\beq
\omega^{\gamma}=\frac 1 2 \eta^{\sigma\alpha\beta\gamma}u_{\sigma} \omega_{\alpha \beta} \label{eq:antisimm}
\eeq
Eq. (\ref{eq:covderUproj1}) can be written in the compact form
\beq
\left(\frac{\mathrm{D}u}{\mathrm{d}\tau_{U}}\right)_{\alpha}=\gamma \left[a(u)_{\alpha}+\theta_{\alpha\beta}v^{\beta}+\left(w  \times _{u} v \right)_{\alpha} \right] \label{eq:covderUproj2}
\eeq
where $v,\omega \in \Sigma$. In fact, using the volume 3-form \cite{Jantzen:1996au}
\beq
\eta_{\alpha\beta\gamma}=u^{\delta}\eta_{\delta\alpha\beta\gamma} \label{eq:volume3form}
\eeq
we may define the spatial cross product of two (spatial) vectors $X,Y$
\beq
\left(X \times _{u}Y\right)^{\alpha} = \eta^{\alpha}_{\, \, \beta \gamma}X^{\beta}Y^{\gamma}
\eeq
which we used in Eq. (\ref{eq:covderUproj2}).

As we are going to show in next Section, the  properties of the congruence play an important role in the description of the dynamics, and the intuitive motivations are simple. In fact, they describe how the reference fluid is moving in spacetime: this is analogous to what happens in Newtonian physics, when we choose an arbitrary moving reference frame to describe dynamics, resulting in inertial effects determined by how the frame moves relative to an inertial one.  The same principle applies to the relativistic frame defined by the congruence: the motion of particles of the reference fluid influences the description of dynamics within it.

\section{Relative Description of Dynamics}\label{sec:reldesc}

In this Section, we focus on the description of the dynamics of a particle\footnote{To fix ideas, we refer to massive particles, however the same approach can be used for massless ones such as photons, using the splitting of their momentum introduced in Section \ref{sec:kin}. } whose momentum is $P=m\,U$ as seen by the reference frame defined by the congruence $\mathcal C_{u}$.  If the particle is freely moving, i.e. if it experiences only gravitational interactions determined by the spacetime metric $g_{\alpha\beta}$, it follows geodesics and its 4-acceleration is null; on the contrary, if it interacts with non-gravitational fields, it possesses a non-null 4-acceleration.

Let $a(U)=\nabla_{U}U$ be the 4-acceleration, and $F(U)$ the 4-vector field representing the ({non-gravitational}) forces acting upon the particle. We set $f(U)=\frac{F(U)}{m}$, so that we have $a(U)=f(U)$. Then, we have two possibilities:
\begin{eqnarray}
f(U) & = & 0 \rightarrow \mathrm{geodesic \  motion} \label{eq:geomot1} \\
f(U) & \neq & 0 \rightarrow \mathrm{non-geodesic \  motion} \label{eq:geomot2}
\end{eqnarray}	

Let us start from the geodesic case. We use Eq. (\ref{eq:4mom7}) to write the 4-momentum in the form\footnote{Here, for the sake of clarity, we simplify the notation and do not explicitly write the dependence on $(U,u)$; for instance we will simply write $P$ instead of $P(U,u)$, and so on.} $P=E\,u+p$, or equivalently
\beq
mU=m\gamma\left(u+v\right). \label{eq:dingeo1}
\eeq
Then, we write the geodesic equation $\nabla_{U}U=0$ in the form 
\beq
\DtauU{U}=0, \label{eq:dingeo2}
\eeq
where $\tau_{U}$ is the proper time measured along the world-line of particle.  Substituting Eq. (\ref{eq:dingeo1}) in (\ref{eq:dingeo2}), we obtain
\beq
\DtauU{\gamma}\,u +\gamma \DtauU{u}+{\DtauU{\left(\gamma v\right)}}=0. \label{eq:relmot1}
\eeq

We notice that the second term in the lhs of Eqs. (\ref{eq:relmot1})  was calculated in Eq. (\ref{eq:covderUproj2}). We are going to focus on its meaning below.  Now, by projecting Eq. (\ref{eq:relmot1}) along $u$, taking into account that $u\cdot u=-1 \rightarrow u\cdot \DtauU{u}=0$, we obtain
\beq
\DtauU{\gamma}={\DtauU{\left(\gamma v\right)}} \cdot u. \label{eq:relmot3}
\eeq

Using Eq. (\ref{eq:relmot3}) in Eq. (\ref{eq:relmot1}) we write
\beq
{\DtauU{\left(\gamma v\right)}}+\left[{\DtauU{\left(\gamma v\right)}} \cdot u \right] u=-\gamma\DtauU{u}. \label{eq:relmot4}
\eeq
Finally, multiplying by the (invariant) mass $m$, we get
\beq
\DtauU{p}+\DtauU{p} \cdot u\, u=-m\gamma \DtauU u. \label{eq:relmot5}
\eeq
Indeed, the lhs of Eq. (\ref{eq:relmot5}) is the space projection of $\displaystyle \DtauU{p}$; in other words
\beq
P(u)\left[ \DtauU{p} \right] = \DtauU{p}+\DtauU{p} \cdot u\, u \doteq \DtauO{p}, \label{eq:relmot51}
\eeq
where we introduced the \textit{transverse derivative operator} $\displaystyle \DtauO{}$.

In summary, we may write
\beq
\DtauO{p}=-m\gamma \DtauU u ,\label{eq:relmot6}
\eeq
Or, equivalently, using the relative time (\ref{eq:reltime3})
\beq
\DTO p =-m \DtauU u. \label{eq:relmot7}
\eeq

The rhs in latter equation depends on the properties of the congruence, i.e. on its acceleration, expansion or vorticity, according to  Eq. (\ref{eq:covderUproj2}), and on the gravitational field. Accordingly, in the relativistic reference frame defined by the congruence $\mathcal C_{u}$, it  plays the role of a gravito-inertial force. So, setting $\displaystyle F^{G} \doteq -m \Dtau u$\footnote{Notice that $F^{G} \in \Sigma$.}, the equation governing the relative dynamics turns out to be
\beq
\DTO p=F^{G}. \label{eq:relmot8}
\eeq
From Eq. (\ref{eq:relmot3}), since $v\cdot u=0$, we obtain
\beq
{ \dtau{\left(m\gamma \right)}}=-p \cdot \Dtau u, \label{eq:relmot9}
\eeq
or, using again the relative time
\beq
{ \dT{\left(m\gamma\right)}}=v\cdot F^{G}.
\eeq

We remark that in this formalism a natural analogy between these gravito-inertial forces and the electromagnetic ones arises: this is the so called \textit{gravito-electromagnetic analogy}. In this case, the acceleration plays the role of the electric field, so it is called \textit{gravito-electric} field, while vorticity plays the role of the magnetic field, so it is referred to as the \textit{gravito-magnetic field} (see e.g. \citet{Ruggiero:2023ker,Jantzen:1996au,Jantzen:1992rg} and references therein).

Since $E=m\gamma$, the relative dynamics is described by the following equations
\begin{eqnarray}
\DTO p &=& F^{G}, \label{eq:newton1} \\
\dT{E}&=&v\cdot F^{G}, \label{eq:newton2}
\end{eqnarray}
which, in the reference frame defined by $\mathcal C_{u}$, manifestly have a Newtonian form. 

As for the {non-geodesic} case (\ref{eq:geomot2}), things are very similar: it is sufficient to add the projection of the non-Newtonian forces $F^{NG}=P(u)F(U)$ to the geodesic case:

\begin{eqnarray}
\DTO p &=& F^{G}+F^{NG,} \label{eq:newton3} \\
\DT{E}&=&v\cdot \left(F^{G}+F^{NG}\right). \label{eq:newton4}
\end{eqnarray}

In summary, we have demonstrated that applying the splitting technique to the equation of motion enables us to describe the dynamics of test particles in terms of effective Newtonian forces. Specifically, in the geodesic case, these forces originate from the properties of the congruence and the background spacetime. Non-gravitational forces add to  the gravito-inertial ones for non-geodesic motions.

Before ending this Section, we add a remark on the meaning of the transverse derivative defined in (\ref{eq:relmot51}). To this end, we remember the Fermi-Walker derivative of a vector field $w$ along the integral curves of $u$ is defined by \cite{synge1960relativity}
\beq
\DLFW{w}=\DL{w}-\left(w\cdot  \DL{u}\right)u+\left(w\cdot u\right)\DL{u}. \label{eq:fw1}
\eeq
where $\lambda$ is the parameter along the curves. In particular, if $w\cdot u=0$, which in our case means that $w \in \Sigma$, we obtain
\beq
\DLFW{w}=\DL{w}-\left(w\cdot  \DL{u}\right)u. \label{eq:fw10}
\eeq

Now, from Eq. (\ref{eq:relmot51}) we have

\beq
\DtauO{p}= \DtauU{p}+\left(\DtauU{p} \cdot u\right) u = \DtauU{p}-\left({p} \cdot \DtauU u\right) u,
\eeq
since $p\cdot u=0$, where  $p=m\gamma v$. So, by comparison with Eq. (\ref{eq:fw10}) we deduce that the transverse derivative coincides with the Fermi-Walker derivative (for space vectors): actually, the relation between the transverse derivative (introduced by \citet{cattaneo1959derivation} and also known as \textit{constrained derivative}) was already emphasized by {\citet{ferrarese}}.

\section{Discussion and Final Remarks}\label{sec:remarks}

We focused on defining  a reference frame in curved spacetime or for observers arbitrarily moving in the flat spacetime of SR, using the approach based on the congruence of world-lines $\mathcal C_{u}$. Physically, this congruence represents the spacetime evolution of the particles constituting the laboratory where the measurement process occurs. Given that, under ordinary conditions, we can assume the laboratory is small enough compared to the length scale of the gravitational field, the congruence is practically identified by the world-line of the observer.

Along this world-line, the introduction of the time  (\ref{eq:defT1}) and  space (\ref{eq:defP1}) projectors {allows us} to obtain, from 4-dimensional entities, space-like quantities and time-like quantities that have an operational meaning in terms of measurements; in doing so, the local Minkowski structure of spacetime naturally emerges.  Accordingly, in Section \ref{sec:kin} we showed that it is possible to describe the motion of test particles  in complete analogy to what happens  in inertial reference frames in SR. This is accomplished in terms of relative quantities, whose essence lies in being ``relative'' to the observer in question. This implies that all observers describe motion on an equal basis, employing the same equations, even though the results of their measurements may differ. We believe this is a beautiful illustration of the principle of relativity.

For the sake of simplicity, we have confined ourselves to the case of relativistic kinematics. However, it can be shown that the projection technique can be systematically applied to formulate the measurement process within a general relativistic framework in other contexts as well, such as in the presence of electromagnetic fields or in the study of the physical properties of fluids (see e.g. \citet{de2010classical} and references therein).

As a further example of the application of this approach, it is important to emphasize the role of the spatial metric defined by (\ref{eq:defgammaij}); the latter, in fact represents the  space metric in $\Sigma$, the 3-dimensional local rest space of the observer. Space measurements are performed using $\gamma_{\alpha\beta}$ {which corresponds to saying} that distances are measured using the radar method \cite{landau2013classical,Ruggiero:2003jb}. For instance,  the line element
\beq
ds^{2}=-\left(1-{\Omega^{2}r^{2}}{} \right)dt^{2}+2{\Omega r^{2}}{}dtd\varphi +dr^{2}+r^{2}d\varphi^{2}+dz^{2} \label{eq:conrotmink}
\eeq
describes a  frame rotating with (constant) rotation rate $\Omega$, in Minkowski spacetime \cite{Rizzi:2002sk}. The congruence defined by
$\displaystyle u=\frac{1}{\sqrt{1-\Omega^{2}r^{2}}}\partial_{t}$ describes observers at rest in the rotating frame, and the coordinates are adapted to this congruence. In this case, the spatial metric gives
\beq
d\sigma^{2}=\gamma_{ij}dx^{i}dx^{j}=dr^{2}+\frac{r^{2}d\varphi^{2}}{\sqrt{1-\Omega^{2}r^{2}}}+dz^{2}, \label{eq:ehr}
\eeq
which shows that the geometry of $\Sigma$ is not Euclidean: this is the root of the well known Ehrenfest's paradox (see \citet{Rizzi:2002sk} and references therein).

As we know, free particles move along geodesics, while those interacting with other fields have an acceleration: this is the standard 4-dimensional description of test particles dynamics. However, {the situation is more subtle  if we want to provide a description of dynamics relative to the given congruence. In fact, the projection of the geodesic equation highlights the role of the properties of the congruence which, in analogy with the classical case, produce gravito-inertial forces}. As a consequence, even if a particle is moving along a geodesic, in the observer's reference frame it appears as if it were acted upon by forces, which depends on the acceleration, rotation and expansion of the congruence: only when all of these quantities are null, then the particle looks free in the observer's frame. The 4-dimensional formulation of dynamics is the same in all reference frames, as for  the principle of General Relativity: accordingly, all reference frames are equal. However, according to the relative formulation, we see that in some of them the description is simpler.

Let us focus on a very relevant implication of the relative description of dynamics. Schwarzschild spacetime  is a solution of Einstein's equations in vacuum for a spherically symmetric source (see e.g. \citet{rindler2012essential}); its weak-field limit is suitable to describe the gravitational field of the Earth {and the line element is}
\begin{align}
    ds^2 =- \left(1-\frac{2M}{ r}\right) dt^2+\left(1+\frac{2M}{ r}\right) dr^2+ r^2 \left(d\theta^2+r^2 \sin^2\theta  d\phi^2\right), \label{eq:cons}
\end{align}
where $M$ is the mass of the source.

We consider the observers  defined by the congruence $\displaystyle u=\frac{1}{\sqrt{1-\frac{2M}{ r}}}\partial_{t}$: they are \textit{static} since they are at rest with respect to the coordinates $(r,\theta,\phi)$; so these coordinates are adapted according to the definition given above. If we neglect Earth rotation, this congruence defines observers at rest on the Earth surface: they are not geodesic observers, since  external forces are needed to make them stay fixed with respect to the coordinates. For instance, the contact force on the Earth surface balances the effect of the gravitational field. 

The gravito-inertial forces appearing in Eq. (\ref{eq:relmot8}) and defined by Eq. (\ref{eq:covderUproj1}) can be calculated in this case, and we obtain that $\displaystyle a(u)_{\alpha}=\frac{M}{r^{2}} \delta_{\alpha}^{r}\partial_{r}$, while $k_{\alpha\beta}=0$. Accordingly, in this frame the presence of the gravitational field is manifested by the {non-inertial} feature of the congruence:  the gravitational field appears as the acceleration  field and
 the description of the motion of free particles is given in terms of the action of this gravito-inertial force. In other words, we observe that the Newtonian gravitational force naturally emerges due to the specific choice of the congruence: it represents an inertial force arising because the observers are not in free fall. Gravitational and inertial forces are treated equally, {in accordance} with the principle of equivalence.
 
Let us study another application of the formalism that we introduced, again in the  Schwarzschild spacetime. We consider a null particle, i.e. an electromagnetic signal, propagating  with 4-momentum $P$ and  two observers at rest at different locations, and their world-lines are defined by the tangent vectors 
\beq
u_{A}=\frac{1}{\sqrt{1-\frac{2M}{ r_{A}}}}\partial_{t}, \quad u_{B}=\frac{1}{\sqrt{1-\frac{2M}{ r_{B}}}}\partial_{t}. \label{eq:Cuudef}
\eeq
Hence, the observers are at rest at $r=r_{A}, r=r_{B}$ respectively. We can calculate the energy which these observers attribute to the signal: this energy is proportional to the frequency of the electromagnetic signal as seen by the observers.
So, from Eq. (\ref{eq:4mom5}), we can write 
\beq
E_{A}=-u_{A}^{0}P_{0},\quad E_{B}=-u_{B}^{0}P_{0}. \label{eq:EAEB}
\eeq
Accordingly, we get
\beq
\frac{E_{A}}{E_{B}}=\frac{u^{0}_{A}}{u^{0}_{B}}=\frac{\sqrt{1-\frac{2M}{ r_{B}}}}{\sqrt{1-\frac{2M}{ r_{A}}}},  \label{eq:EAEB1}
\eeq 
which is indeed the relation for the gravitational redshift \cite{rindler2012essential}. \\

In conclusion, we demonstrated how the splitting approach allows us to translate the formulation of physics laws, conducted in the 4-dimensional spacetime of General Relativity, into actual measurement processes occurring in space and time within a sufficiently small neighborhood along the observer's world-line. Consistent with the saying ``physics is simple when analysed locally'', we emphasized that physics is locally Lorentzian everywhere. Additionally, we focused on interpreting forces within the local reference frame, noting that the Newtonian gravitational force is a manifestation of the non-geodesic character of the reference frame. We believe that this approach provides an effective means to elucidate the essence of relativistic physics while establishing clear connections with familiar concepts rooted in classical physics.

\bibliography{splitting_draft}

\end{document}